\documentclass{PoS}

\PoS{PoS(LAT2005)156}

\usepackage{fleqn,epsf}

\def\mpi2{m_\pi^2}
\def\mK2{m_K^2}

\newcommand{\bea}{\begin{eqnarray}}
\newcommand{\eea}{\end{eqnarray}}
\newcommand{\be}{\begin{equation}}
\newcommand{\ee}{\end{equation}}

\def\lvec#1{\setbox0=\hbox{$#1$}
    \setbox1=\hbox{$\scriptstyle\leftarrow$}
    #1\kern-\wd0\smash{
    \raise\ht0\hbox{$\raise1pt\hbox{$\scriptstyle\leftarrow$}$}}
    \kern-\wd1\kern\wd0}
\def\rvec#1{\setbox0=\hbox{$#1$}
    \setbox1=\hbox{$\scriptstyle\rightarrow$}
    #1\kern-\wd0\smash{
    \raise\ht0\hbox{$\raise1pt\hbox{$\scriptstyle\rightarrow$}$}}
    \kern-\wd1\kern\wd0}

\newcommand{\VEV}[1]{\left\langle #1\right\rangle}
\newcommand{\Tr}{\mbox{Tr}}
\newcommand{\nn}{\nonumber}

\title{The Equation of State for QCD with 2+1 Flavors of Quarks}

\ShortTitle{The Equation of State for QCD with 2+1 Flavors of Quarks}

\author{C.~Bernard$^a$,  T.~Burch$^b$, C.~DeTar$^c$, Steven~Gottlieb$^d$, 
        U.~M.~Heller$^e$, J.~E.~Hetrick$^f$,
        \speaker{L.~Levkova}$^{\,\,d}$, 
        F.~Maresca$^c$,
        D.~B.~Renner$^g$, 
        R.~Sugar$^h$, D.~Toussaint$^g$\\
        \llap{$^a$}Physics Department, Washington University, St. Louis, MO 63130, USA\\
        \llap{$^b$}Institut f\"ur Theoretische Physik, Universit\"at Regensburg, D-93040 Regensburg, Germany\\
        \llap{$^c$}Physics Department, University of Utah, Salt Lake City, UT 84112, USA\\
	\llap{$^d$}Physics Department, Indiana University, Bloomington, IN 47405, USA\\
        \llap{$^e$}American Physical Society, One Research Road, Box 9000, Ridge, NY 11961-9000, USA\\
	\llap{$^f$}Physics Department, University of the Pacific, Stockton, CA 95211, USA\\
        \llap{$^g$}Physics Department, University of Arizona, Tucson, AZ 85721, USA\\
        \llap{$^h$}Physics Department, University of California, Santa Barbara, CA 93106, USA\\
         E-mail: \email{cb@lump.wustl.edu}, 
                \email{tommy.burch@physik.uni-regensburg.de}, 
                \email{detar@physics.utah.edu},
                \email{sg@indiana.edu}, 
                \email{heller@aps.org}, 
                \email{jhetrick@uop.edu},
                \email{llevkova@indiana.edu}, 
                \email{marescaf@maths.tcd.ie}, 
                \email{dru@physics.arizona.edu},
                \email{sugar@physics.ucsb.edu},
                \email{doug@physics.arizona.edu}}

\abstract{
We report results for the interaction measure, pressure and energy density
for nonzero temperature QCD with 2+1 flavors of
improved staggered quarks. In our simulations we use a Symanzik improved
gauge action and the Asqtad $O(a^2)$ improved staggered quark action
for lattices with temporal extent $N_t=4$ and 6. The heavy quark mass $m_s$
is fixed at approximately  the physical strange quark mass and the
two degenerate light quarks have masses $m_{ud} =0.1m_s$ or $0.2m_s$.
The calculation of the thermodynamic observables employs the integral
method where energy density and pressure are obtained by integration
over the interaction measure.}

\FullConference{XXIIIrd International Symposium on Lattice Field Theory\\
                 25-30 July 2005\\
                 Trinity College, Dublin, Ireland}

\begin{document}
\section{Introduction}

The equation of state (EOS) is important for phenomenological models of quark-gluon
plasma formation and decay, which is currently under experimental study at
RHIC and elsewhere.
We have determined the EOS with the Asqtad quark action \cite{asq} for 2+1 flavors, 
combined with a one-loop Symanzik improved gauge action \cite{sym}. The Asqtad action 
is well suited for high temperature studies since it has excellent scaling properties
and much better dispersion relations in the free case than the standard Wilson
or staggered actions, which means decreased lattice artifacts above the transition.

Our nonzero temperature studies are at $N_t=4$ and 6.
Even with Asqtad improvement, for  $T \leq T_c$, an $N_t = 4$ lattice has
a badly split pion taste multiplet with some members heavier than
the kaon, which makes for questionable strange quark physics.  At
$N_t = 6$ the taste-splitting is about half as large. One of our goals was to determine
to what extent the increase in $N_t$ from 4 to 6 influences the EOS.

\section{Action}

The fermion part of the action we use is effectively written as:
\begin{small}
\be
S_f = -\sum_f (n_f/4) \Tr\ln[M(am_f,U,u_0)],
\ee
\end{small}
where $M(am_f,U,u_0)$ is the fermion matrix corresponding to the Asqtad
2+1 flavor staggered action.
The gauge part is defined as:
\begin{small}
\begin{equation}
  S_g = \beta \sum_{x,\mu<\nu} (1 - P_{\mu\nu}) 
        + \beta_{\rm rt} \sum_{x,\mu<\nu} (1 - R_{\mu\nu})
        +\beta_{\rm ch} \sum_{x,\mu<\nu<\sigma} (1 - C_{\mu\nu\sigma}).
\end{equation}
\end{small}
The gauge couplings above are
$\beta = 10/g^2$, $\beta_{\rm rt} = -\beta (1 + 0.4805 \alpha_s)/(20u_0^2)$,
$\beta_{\rm ch} = - 0.03325 \beta\alpha_s/u_0^2$,
with $\alpha_s = -4 \ln(u_0)/3.0684$ and $u_0 = \VEV{P}^{1/4}$.

For our simulations we use the dynamical R-algorithm \cite{Ralg} with step-size equal to the smaller of 0.02 and 
$2am_{ud}/3$. 
Our aim is  
to generate zero  and nonzero temperature ensembles of lattices with
action parameters chosen so that a constant physics trajectory ($m_\pi/m_\rho = const$)
is approximated. Along the trajectory the heavy quark mass is fixed close to the strange quark mass.
We work with two such trajectories:
$m_{ud}=0.2m_s$ ($m_\pi/m_\rho \approx 0.4$) and 
$m_{ud}=0.1m_s$ ($m_\pi/m_\rho \approx 0.3$).

\section{Parameterization of the Constant Physics Trajectories and Run Parameters} 

 Our trajectories are intended to approximate constant zero
      temperature physics.  The construction of each trajectory begins
      with "anchor points" in $\beta$, where the hadron spectrum has
      been previously studied \cite{had} and the lattice strange quark mass
      has been tuned to approximate the correct strange hadron spectrum.
      We adjusted the value of $a m_{ud}$ at the anchor points to give a
      constant (unphysical) ratio $m_\pi/m_\rho$.  Between these points
      the trajectory is then interpolated, using a one-loop
      renormalization group inspired formula.  That is, we interpolate
      $\ln(a m_s)$ and $\ln(a m_{ud})$ linearly in $\beta$.  Since we have
      three anchor points for the $m_{ud} = 0.2 m_s$ trajectory,
      namely $\beta = 6.467$, 6.76, and 7.092, our interpolation is
      piecewise linear.  For the trajectory $m_{ud} = 0.1 m_s$ we use
      two anchor points at $\beta = 6.458$ and 6.76.
For both trajectories, for values of $\beta$ out of the interpolation intervals, the
parameterization formulas are used as extrapolations appropriately. 
 The run parameters of the two trajectories at different $N_t$
are summarized in Tables~1, 2 and 3. 
\begin{table}
\begin{center}
\begin{footnotesize}
\begin{tabular}{|l|l|l|l|l|l|}   \hline\hline
 $\beta$ & $am_{ud}$ & $am_s$ & $u_0$ & $V_{T\neq0}$ & $V_{T=0}$ \\ \hline\hline

$^\star$6.300 & 0.0225 & 0.1089 & 0.8455& $12^3\times6$& $12^4$\\\hline

$^\star$6.350 & 0.0206 & 0.1001 &0.8486&$12^3\times6$&$12^4$ \\\hline
 6.400 &0.01886 &0.0919 & 0.8512&$12^3\times6$&  \\\hline
 6.433 &0.01780 &0.0870 &0.8530&$12^3\times6$& \\\hline
$^\star$6.467 &0.01676 &0.0821 &0.8549&$16^3\times6$&$16^3\times48$ \\\hline
 6.500 &0.01580 &0.0776 &0.8568& $12^3\times6$&\\\hline
$^\star$6.525 &0.01510 &0.0744 &0.8580&$12^3\times6$& $12^4$ \\\hline
 6.550 &0.01450 &0.0713 &0.8592 &$12^3\times6$&\\\hline
$^\star$6.575 &0.01390 &0.0684 &0.8603 &$12^3\times6$& $16^4$\\\hline
 6.600 &0.01330 &0.0655 &0.8614&$12^3\times6$& \\\hline
$^\star$6.650 &0.01210 &0.0602 &0.8634&$12^3\times6$& $20^4$\\\hline
 6.700 &0.01110 &0.0553 &0.8655 &$12^3\times6$&\\\hline
$^\star$6.760 &0.01000 &0.0500 &0.8677& $20^3\times6$&$20^3\times64$\\\hline
 7.092 &0.00673 &0.0310 &0.8781& $12^3\times6$&\\\hline
 7.090 &0.00620 &0.0310 &0.8782& & $28^3\times96$\\\hline
\end{tabular}
\end{footnotesize}
\end{center}
\caption{Run parameters of the trajectory with $m_{ud}=0.2m_s$ at $N_t=6$. The asterisk indicates parameter sets 
for which both zero  and nonzero temperature runs were performed.}
\end{table}

\begin{table}
\begin{center}
\begin{footnotesize}
\begin{tabular}{|l|l|l|l|l|l|}   \hline\hline
 $\beta$ & $am_{ud}$ & $am_s$ & $u_0$ & $V_{T\neq0}$ & $V_{T=0}$ \\ \hline\hline
$^\star$6.300 & 0.01090 & 0.1092 & 0.8459 &$12^3\times6$ & $12^4$\\\hline
$^\star$6.350 & 0.00996 & 0.0996 & 0.8491 &$12^3\times6$ & $12^4$\\\hline
        6.400 & 0.00909 & 0.0909 & 0.8520 &$12^3\times6$ & \\\hline
$^\star$6.458 & 0.00820 & 0.0820 & 0.8549 &$16^3\times6$ & $12^4$\\\hline
        6.500 & 0.00765 & 0.0765 & 0.8570 &$12^3\times6$ & \\\hline
$^\star$6.550 & 0.00705 & 0.0705 & 0.8593 &$12^3\times6$ & $20^4$\\\hline
        6.600 & 0.00650 & 0.0650 & 0.8616 &$12^3\times6$ & \\\hline
$^\star$6.650 & 0.00599 & 0.0599 & 0.8636 &$12^3\times6$ & $24^4$\\\hline
        6.700 & 0.00552 & 0.0552 & 0.8657 &$12^3\times6$ & \\\hline
$^\star$6.760 & 0.00500 & 0.0500 & 0.8678 &$20^3\times6$ & $24^3\times 64$\\\hline
$^\star$7.080 & 0.00310 & 0.0310 & 0.8779 &$18^3\times6$ & $40^3\times 96$\\\hline
\end{tabular}
\end{footnotesize}
\end{center}
\caption{Run parameters of the trajectory with $m_{ud}=0.1m_s$ at $N_t=6$.  The asterisk indicates parameter sets 
for which both zero  and nonzero temperature runs were performed.}
\end{table}

\begin{table}
\begin{center}
\begin{footnotesize}
\begin{tabular}{|l|l|l|l|l|l|}   \hline\hline
 $\beta$ & $am_{ud}$ & $am_s$ & $u_0$ & $V_{T\neq0}$ & $V_{T=0}$ \\ \hline\hline
$^\star$6.000 & 0.01980 & 0.1976 & 0.8250 &$12^3\times4$ & $12^4$\\\hline
$^\star$6.050 & 0.01780 & 0.1783 & 0.8282 &$12^3\times4$ &$12^4$\\\hline
             6.075 & 0.01690 & 0.1695 & 0.8301 &$12^3\times4$ &\\\hline
$^\star$6.100 & 0.01610 & 0.1611 & 0.8320 &$12^3\times4$ &$12^4$\\\hline
             6.125 & 0.01530 & 0.1533 & 0.8338 &$12^3\times4$ &\\\hline
$^\star$6.150 & 0.01460 & 0.1458 & 0.8356 &$12^3\times4$ &$12^4$\\\hline
             6.175 & 0.01390 & 0.1388 & 0.8374 &$12^3\times4$ &\\\hline
$^\star$6.200 & 0.01320 & 0.1322 & 0.8391 &$12^3\times4$ &$12^4$\\\hline
             6.225 & 0.01260 & 0.1260 & 0.8407 &$12^3\times4$ &\\\hline
$^\star$6.250 & 0.01200 & 0.1201 & 0.8424 &$12^3\times4$ &$12^4$\\\hline
             6.275 & 0.01140 & 0.1145 & 0.8442 &$12^3\times4$ &\\\hline
$^\star$6.300 & 0.01090 & 0.1092 & 0.8459 &$12^3\times4$ &$12^4$\\\hline
$^\star$6.350 & 0.00996 & 0.0996 & 0.8491 &$12^3\times4$ &$12^4$\\\hline
             6.400 & 0.00909 & 0.0909 & 0.8520 &$12^3\times4$ &\\\hline
$^\star$6.458 & 0.00820 & 0.0820 & 0.8549 &$12^3\times4$ &$12^4$\\\hline
             6.500 & 0.00765 & 0.0765 & 0.8570 &$12^3\times4$ &\\\hline
$^\star$6.550 & 0.00705 & 0.0705 & 0.8593 &$12^3\times4$ &$20^4$\\\hline
             6.600 & 0.00650 & 0.0650 & 0.8616 &$12^3\times4$ &\\\hline
$^\star$6.650 & 0.00599 & 0.0599 & 0.8636 &$12^3\times4$ &$24^4$\\\hline
             6.700 & 0.00552 & 0.0552 & 0.8657 &$12^3\times4$ &\\\hline
$^\star$6.760 & 0.00500 & 0.0500 & 0.8678 &$12^3\times4$ &$24^3\times 64$\\\hline
\end{tabular}
\end{footnotesize}
\end{center}
\caption{Run parameters of the trajectory with $m_{ud}=0.1m_s$ at $N_t=4$.  The asterisk indicates parameter sets 
for which both zero  and nonzero temperature runs were performed.}
\end{table}

\section{Integral Method for the EOS derivation}

To derive the analytic form of the EOS we employ the integral method \cite{int}. 
We start from the thermodynamic identities:
\begin{small}
\be
  \varepsilon V = - \left.\frac{\partial \ln Z}{\partial(1/T)}\right|_V,\hspace{0.7cm}
  \frac{p}{T} = \left.\frac{\partial \ln Z}{\partial V}\right|_T \approx \frac{\ln Z}{V},\hspace{0.7cm}
   I = \varepsilon - 3p = -\frac{T}{V} \frac{\partial \ln Z}{\partial \ln a},
\ee\end{small}
where the derivative with respect to $\ln a$ is taken at
constant $m_\pi/m_\rho$ and the partition function is $Z$. 
Using the above identities  and the explicit form of $Z$ we obtain:
\begin{small}
\bea
Ia^4&=&-6\frac{d \beta} {d \ln a}\Delta\VEV{P}
       - 12\frac{d \beta_{\rm rt}} {d \ln a}\Delta\VEV{R}
       - 16\frac{d \beta_{\rm ch}} {d \ln a}\Delta\VEV{C}\\
    &-& \sum_f \frac{n_f}{4}\left[\frac{d (m_f a)}{d \ln a}
           \Delta\VEV{\bar \psi \psi}_f
       + \frac{d u_0}{d \ln a}
       \Delta\VEV{\bar\psi\, \frac{d M}{d u_0}\, \psi}_f\right],\nn\\
  pa^4 &=& \int_{\ln a_0}^{\ln a}\left\{
       6\frac{d \beta} {d \ln a}\Delta\VEV{P}
       + 12\frac{d \beta_{\rm rt}} {d \ln a}\Delta\VEV{R}
       + 16\frac{d \beta_{\rm ch}} {d \ln a}\Delta \VEV{C}\right.\\
       &+& \left.\sum_f \frac{n_f}{4}\left[\frac{d (m_f a)}{d \ln a}
           \Delta\VEV{\bar \psi \psi}_f
       + \frac{d u_0}{d \ln a}
       \Delta\VEV{\bar\psi\, \frac{d M}{d u_0} \,
         \psi}_f\right]\right\} \, d\ln a^\prime,\nn \\
\varepsilon a^4 &=& (I+3p)a^4.
\eea\end{small}
In the above expressions the symbol $\Delta$ in front of the lattice observables stands for the difference between
the values of those observables at nonzero and zero temperatures.
In the pressure expression, the lower integration endpoint $\ln a_0$ is set where the
zero temperature subtracted value of $Ia^4$ is zero within errors at coarse lattice spacings.
To calculate the EOS, in addition to the lattice gluonic and fermionic observables in the above analytic forms,
we need to calculate the derivatives
$d \beta_{\rm pl}/d \ln a$,
$d \beta_{\rm rt}/d \ln a$,
$d \beta_{\rm ch}/d \ln a$,
$d (m_f a)/d \ln a$ and
$d u_0/d \ln a$.
For this purpose we take derivatives of
the $\ln(am_{ud})$ and $\ln(am_{s})$ trajectory parameterizations,
polynomial fits to $u_0(\beta)$ for both trajectories, and the updated version of the $\ln(r_1/a)$
 fitting formula in \cite{had} shown below:
\begin{small}
\bea
\ln(r_{1}/a) &=& C_{00} + C_{10}(\beta-7) + C_{21}(2am_{ud}+am_{s})+ C_{20}(\beta-7)^2,
\eea
\end{small}
where $r_{1}=0.317(7)(3)$ fm, $C_{00} = 1.261(3)$,  $C_{10} = 0.939(8)$,  $C_{21} = -0.86(3)$ 
and  $C_{20} = -0.25(2)$. 
The above fit (4.5) giving the relative lattice scale is based on
measurements of the static quark potential at zero temperature
for a large set of $\beta$ and quark masses.  The absolute scale
is fixed from a determination of the bottomonium spectrum
\cite{bot}. The fit has $\chi^2/DOF \approx 1.1$.

\section{EOS results}

Figure~1 summarizes our results for the EOS. 
The errors on all data points are calculated using the jackknife method, and we ignore insignificant
errors on the derivatives of the bare parameters with respect to the lattice scale discussed 
at the end of the previous section. 
For the points where there is no zero temperature run, local interpolations are
made to calculate the zero temperature corrections to the interaction measure. 
The integration of the interaction measure to obtain the pressure is done using the trapezoid method.

The comparison between the $N_t=4$ and 6 cases 
for the $m_{ud}=0.1m_{s}$ trajectory shows that there is not a significant difference between them, 
except in the interaction measure near the transition region.
The EOS results from the two different physics trajectories are very similar and for the 
temperature interval we studied, the deviation from the 3 flavor Stefan--Boltzmann values are large.

In the temperature region where we have data,
we consider our pressure results to be in general agreement to a previous p4-action 
$N_t=4$ calculation \cite{karsch}.

\begin{figure}[t]
\epsfxsize=\hsize
\begin{center}
\epsfbox{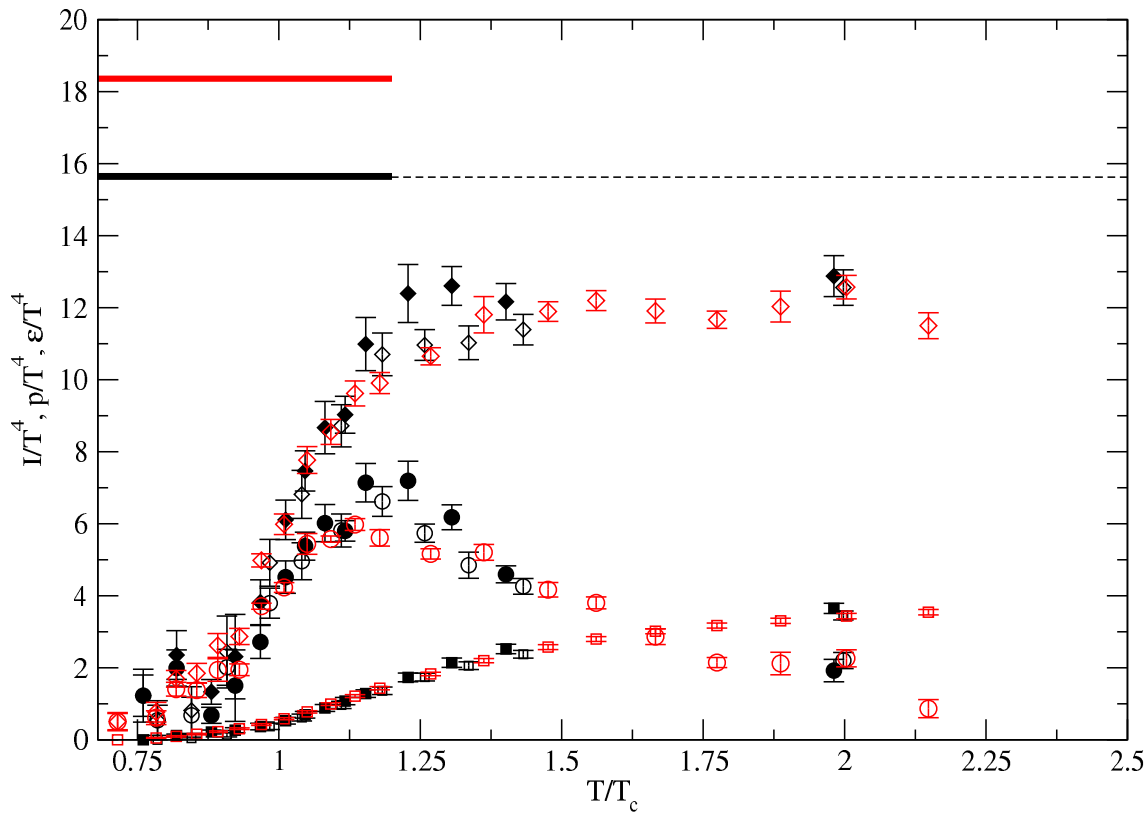}
\end{center}\vspace{-0.7cm}
\caption{The EOS with 2+1 dynamical quark flavors for both constant physics trajectories. The energy
density, pressure and the interaction measure symbols are diamonds, squares and circles, respectively. 
The data at $N_t=6$ is in black and at $N_t=4$ in red. 
All data with filled symbols belong to the $m_{ud} =0.2m_s$ trajectory and 
the empty symbols belong to the $m_{ud} =0.1m_s$ trajectory. The Stefan--Boltzmann
lattice limits for 3 flavors of quarks at $N_t=4$ and 6 are represented by solid 
lines -- red and black, respectively.
The black dashed
line is the continuum Stefan--Boltzmann limit for 3 flavors of quarks.}
\end{figure}

\section{Conclusions}

We have calculated the EOS for 2+1 dynamical flavors of improved staggered quarks
  ($m_{ud}/m_{s}=0.1$ and 0.2) along  trajectories of constant physics, at
$N_t=4$ and 6, where the latter is the first result of its kind.
Our results show that the $N_t=4$ and $N_t=6$ results are quite similar
     except in the crossover region where the interaction
     measure is a bit higher on the finer $N_t=6$ lattice. We also do not see significant differences
between  the EOS results from the two physics trajectories. We find large deviations from the
3 flavor Stefan--Boltzmann limit. 
Our results are comparable with previous calculations \cite{karsch}. 
\begin{center}\bf{ Acknowledgments}\\
\end{center}

This work was supported by the US DOE and NSF. 
Computations were performed at CHPC (Utah), FNAL, FSU, IU, NCSA and UCSB.

\end{document}